\documentclass[11pt,pra,aps,onecolumn,superscriptaddress,floatfix,notitlepage,nofootinbib]{revtex4-2}
\usepackage[latin1]{inputenc}
\usepackage{hyperref}
\hypersetup{
    colorlinks=true,
    linkcolor=blue,
    filecolor=magenta,      
    urlcolor=cyan,
    pdftitle={Complementary Noise in Radiation Fields},
    pdfpagemode=FullScreen,
    }
\usepackage{mathrsfs,dsfont}
\usepackage{amsmath}
\usepackage{amsfonts,natbib}
\usepackage{amssymb}
\usepackage{bm}
\usepackage[pdftex]{graphicx}

\usepackage{array}

\usepackage{textpos}
\usepackage{bbold}
\usepackage{adjustbox}

\usepackage{textpos}
\usepackage{bbold}
\usepackage{adjustbox}

\usepackage[normalem]{ulem}

\usepackage{amsmath}
\usepackage{graphicx}

\begin{document}
	\title{Probing Quantum Structure in Gravitational Radiation\footnote{Essay for the 2025 Gravitation Research Foundation competition.  First Prize.}} 
\author{Sreenath K. Manikandan}
\affiliation{Nordita, Stockholm University and KTH Royal Institute of Technology, Hannes Alfv\'{e}ns v\"{a}g 12, SE-106 91 Stockholm, Sweden}
\author{Frank Wilczek}
\affiliation{Department of Physics, Arizona State University, Tempe, Arizona 25287, USA}
\affiliation{T. D. Lee Institute, Shanghai 201210, China}
\affiliation{Wilczek Quantum Center and Department of Physics and Astronomy, Shanghai Jiao Tong University, Shanghai 200240, China }
\affiliation{Department of Physics, Stockholm University, AlbaNova University Center, 106 91 Stockholm, Sweden}
\affiliation{Nordita, Stockholm University and KTH Royal Institute of Technology, Hannes Alfv\'{e}ns v\"{a}g 12, SE-106 91 Stockholm, Sweden}
\affiliation{Center for Theoretical Physics, Massachusetts Institute of Technology, Cambridge, Massachusetts 02139, USA}

\begin{abstract}
Gravitational radiation from known astrophysical sources is conventionally treated classically.  This treatment corresponds, implicitly, to the hypothesis that a particular class of quantum-mechanical states - the so-called coherent states - adequately describe the gravitational radiation field.  We propose practicable, quantitative tests of that hypothesis using resonant bar detectors monitored in coincidence with LIGO-style interferometers.    Our tests readily distinguish fields that contain significant thermal components or squeezing.  We identify concrete circumstances in which the classical (i.e., coherent state) hypothesis is likely to fail.  Such failures are of fundamental interest, in that addressing them requires us to treat the gravitational field quantum-mechanically, and they open a new window into the dynamics of gravitational wave sources.  
\end{abstract}

\maketitle

\bigskip

{\it Coherent state hypothesis}

\bigskip

Under the assumption that the world obeys the laws of quantum theory, the widespread treatment of gravitational radiation as a classical field must be based, at least implicitly, on some hypothesis about the quantum state of that radiation.  The most defensible hypothesis of this kind, whose nature and limitations have been analyzed extensively, both theoretically and experimentally, for the corresponding question in electromagnetic radiation (i.e., in quantum optics), is that the classical field approximation is justified when the radiation field can be expressed to a good approximation as a superposition, over independent modes, of coherent states.  

Coherent states, by definition, are right (respectively, left) eigenstates of the lowering operator $a$~(respectively, raising operator $a^\dagger$)~\cite{Sudarshan,Glauber}. 
Since coherent states are eigenstates of the operators that appear in the expansion of the underlying quantum field, the action of the quantum field on such states reduces to multiplication by an ordinary number -- in Dirac's terminology, a c-number.    In this precise sense, coherent states allow one to implement the idea of a ``classical'' fields within the framework of quantum theory. 

Thus it seems appropriate, and will prove fruitful, to sharpen the vague assertion ``Gravitational radiation is classical.''  into the {\it coherent state hypothesis}, that it can be described accurately as a superposition, over independent modes, of coherent states.  Below we will propose concrete experimental tests of that hypothesis.

One can show that c-number sources coupled linearly to the basic field, be it gravitational or electromagnetic, generate radiation fields that satisfy the coherent state hypothesis.  Intuitively, we might expect that robust sources, such as massive astronomical bodies, for which radiation constitutes are a small perturbation, can be modeled as c-numbers.  It is also justified for the macroscopic polarization fields involves in laser action; and laser light has been shown to obey the coherent state hypothesis accurately, at least well above the lasing threshold.   

On the other hand, it is not reasonable to expect that the radiative transitions of atoms or, in the case of Hawking radiation, the short-distance tunneling events, that give the high-frequency part of thermal radiation can be modeled as c-number sources.  And indeed for thermal radiation the coherent state hypothesis fails, as our tests easily reveal. Below we will identify other cases of gravitational radiation where the coherent state hypothesis is likely to fail in a way that is accessible to observation, using our tests.  They arise because the sources are {\it nonlinearly \/} coupled to the gravitational field.

\bigskip

{\it Quantized response of bar detectors} 

\bigskip

Resonant mass or bar detectors for gravitational radiation were pioneered by Joseph Weber.  Weber's idea was to measure resonant acoustic oscillations of a massive cylindrical bar that arise in response to tidal forces exerted by incident gravitational waves.  This strategy failed to achieve the sensitivity required to convincingly identify gravitational waves from astrophysical sources, and it has been superseded by LIGO-style interferometry as a discovery tool within their overlapping frequency ranges.  

Nevertheless there has been continuing interest in bar detectors, and several are presently in operation around the world.  Benefiting from modern developments in computer control, materials fabrication, cryogenics, and other technologies, they have advanced far beyond Weber's pioneering early efforts.  While no single bar detector can at present identify gravitational radiation from astrophysical sources with adequate statistical significance, when several are run in parallel, and sampled in coincidence with interferometric detection (thus limiting the window for noise), they can give confirmatory detection.  Note that because bars are much smaller than the wavelength of the gravitational radiation they resonate with, it is practical to have separated bars that run in parallel.  This could be very helpful for determining the intrinsically probabilistic response of bars, which is characteristic of quantum effects, since it allows sampling based on a single astronomical event.

Recent theoretical work has made it plausible that individual quanta -- i.e., phonons -- excited by gravitation radiation can be detected in realizable bar detectors~\cite{Graviton}. Let us call these detections ``clicks''. The bar detectors in question are not useful for discovering astrophysical sources, because their conversion efficiency is small. It takes a stimulating radiation field that contain as many as $10^{36}$ gravitons to produce a single click.  Fortunately, bar detectors naturally operate in the LIGO band frequencies, and  known sources of gravitational radiation are calculated to be adequate produce a click with probability of order unity.  Thus, observation of single clicks in a bar detector due to gravitational radiation is a realistic aspiration for near-future experiments.  It would demonstrate that gravitational radiation exchanges energy with matter in a quantized fashion and thereby significantly strengthen, through empirical evidence, theoretical arguments that the gravitational field must be quantized. 

\bigskip

{\it Test of the coherent state hypothesis using counting statistics}

\bigskip

Ironically, the fact that bar detectors are on the edge of failure -- that is, of not clicking -- implies they are especially sensitive to fluctuations.   Thus, it is plausible {\it a priori\/} that fluctuations in the response of bar detectors might be a promising window into variable quantum structure within the gravitational radiation they monitor.  

In their most straightforward, stand-alone operating mode bar detectors cleanly sample a discrete spectrum of resonant modes in a digital fashion.  Indeed, each mode can be well modeled as a harmonic oscillator, which might be unexcited, singly excited, doubly excited ... by a given gravitational radiation event.  Alternatively, we can say that the event produced, zero, one, two, ... phonons.  The quantized response of the detector is probabilistic, and the statistics of this response provides our most basic test of the coherent state hypothesis.

Meaningful statistical inferences can be made by using the following interaction Hamiltonian between the acoustic mode of a resonant detector and gravitational radiation~\cite{Acoherence},
\begin{equation}\label{interaction}
    H_{I} \Delta t = \hbar\sqrt{\gamma_0\Delta t}(a^\dagger b+b^\dagger a).
\end{equation}
The mode $b$ is the detector (acoustic mode of the resonant mass detector), the mode $a$ represents the gravitational radiation field in the single mode approximation, and $\gamma_0$ is the spontaneous emission rate of the detector for the field quantum (graviton). The conversion efficiency $\gamma_0\Delta t $ encapsulates the details of the interaction.  The weakness of gravitational interaction is evident from the fact that $\gamma_0\sim 10^{-33}s^{-1}$ for a typical bar detector.   

Since both the resonant gravitational radiation mode -- a mode of the free gravitational field -- and the bar mode are effectively governed by harmonic oscillator dynamics, and the interaction (\ref{interaction}) is a bilinear coupling between them, one can calculate the evolution of the coupled system exactly, at the level of operators, and apply the solution to states of interest, both for the radiation and for the detector.  

For simplicity, let us first consider that the detector begins in its ground state.  Let $P_n$ denote the probability of detecting $n$ phonons in a given event.  For coherent states the of the radiation field mode the statistics of counts follows Poisson statistics: $P_n = e^{-\lambda} \lambda^n/ n!$, where $\lambda$ is the mean count.  This ``maximally classical'' result corresponds to uncorrelated counts, reflecting the absence of (quantum) noise in this case.  A simple consequence, practically accessible for low count rates, is that the ratio
\begin{equation}\label{simple_ratio}
 R \equiv \frac{2P_2P_0}{P_1^2}
\end{equation}
is equal to one: $R_{\rm coherent} =1$.  On the other hand for thermal states (at any temperature) $R_{\rm thermal} = 2$, and for squeezed vacuum states $R_{\rm vacuum \ squeezed} = 2 \, + \, \coth^2(r) $, so that $R_{\rm vacuum \ squeezed} \rightarrow 3$ for strong squeezing.  One can obtain analytic results for a wide variety of non-trivial states that combine thermal, displacement, and squeezing operations.  Significant deviations from $R=1$ are generic. 

It is also informative to consider the global statistics
\begin{eqnarray}
    \bar{n} &=& \sum_{n=0}^{\infty} n P_{n} =\sin^2(\sqrt{\gamma_0\Delta t}) \langle a^\dagger a\rangle\approx \gamma_0\Delta t \langle a^\dagger a\rangle \\
 (\Delta n)^2 &\approx&\bar{n}+(\gamma_0\Delta t)^2 Q \langle a^\dagger a \rangle,\label{eqvar}   
\end{eqnarray}
Here we have introduced Mandel's $Q$ parameter
\begin{eqnarray}
 Q \equiv   \frac{\langle(\Delta \hat{N})^2\rangle -\langle \hat{N}\rangle}{\langle \hat{N}\rangle}
\end{eqnarray}
where $\hat{N} \equiv a^{\dagger}a$. Note that $R = 1 + Q/\langle \hat{N} \rangle$.

Since the requirement for a detectable signal is that $\gamma_0\Delta t \langle \hat{N}\rangle$ is of order unity, and $\gamma_0\Delta t$ is tiny, detectable deviations from the coherent state predictions, in either the ratio test or the global statistics, require in addition that $Q$ is of order $\langle \hat{N} \rangle$. When both these conditions are satisfied, departures from a coherent state becomes visible as excess quantum noise in the resonant detector~\cite{Acoherence}.  As we noted above, both thermal states and highly squeezed vacuum states of appropriate strength satisfy both requirements, and do exhibit observable deviations from coherent states based on the above criteria. In both these cases, and indeed in all cases where $Q >>1 $, the count fluctuations are super-Poissonian.

On the other hand states leading to sub-Poissonian statistics must obey $-1 \leq Q < 0$.  As a practical matter, they cannot be distinguished from coherent states solely on the basis of counting statistics.  Eigenstates of number (Fock states) with large occupancy are sub-Poissonian ($Q=-1$) with $R \approx 1$; they are highly non-classical, but our tests do not distinguish them from coherent states.  The difficulty of observing the reduction in noise below the vacuum noise owing to quantum mechanical effects in gravitational radiation, due the low graviton to click conversion efficiencies has been emphasized recently in ~\cite{Carney1,Carney2}.

One might reasonably hope, however, that lack of fluctuations in number will often be associated with enhanced fluctuations in phase, since phase is conjugate to number.  This serves as extra motivation (if any were needed) to examine the possibility of accessing phase information using bar detectors.

\bigskip

{\it Interferometry at the Detector}

\bigskip

Homodyne and heterodyne operation of detectors are two interferometric methods of accessing information about radiation fields that cannot be obtained from clicks, as considered above.  Both are based on measuring interference between the signal induced by the radiation and the signal induced by a known, observer-controlled local oscillator.  Here will not discuss the technicalities of these methods, but focus directly on the observables they access.  

Homodyne operation allows us to measure the excitation of the detector in a quadrature basis, for example (and without loss of generality) the basis formed by the effective position eigenstates $| x \rangle $, that satisfy
\begin{equation}
x_0 \frac{b+b^\dagger}{\sqrt 2} |x \rangle = x |x \rangle
\end{equation}
We calculate that the variance of $x$ in the {\it detector\/} is related to the variance of the conjugate variable $P=\frac{a-a^\dagger}{i\sqrt{2}}$ in the {\it radiation}, according to 
\begin{equation}
    \langle (\Delta x)^2\rangle_D = x_0^2\bigg[\frac{1}{2}+\sin^2(\sqrt{\gamma_0\Delta t})\bigg(\langle (\Delta P)^2\rangle-\frac{1}{2}\bigg)\bigg].
\end{equation}
where $x_0 \propto \sqrt{\hbar}$ is the effective zero-point length for the detector mode.  

This has a remarkable application to the problem mentioned previously, of differentiating between radiation that generates sub-Poissonian count fluctuations and coherent radiation.
For a coherent radiation state $\langle (\Delta P)^2\rangle =1/2$, so that the fluctuations reduce to detector quantum noise.  But for a number state $|n\rangle$, one has $\langle (\Delta P)^2\rangle =(2 n+1)/2$.   Evidently the excess noise observable from a maximally sub-Poissonian number state, when compared to a coherent state can be substantial.  In fact, if $n$ is large enough so that $\gamma_0\Delta t n\sim O(1)$, then observable deviation from coherent state expectations in the homodyne noise is also of order unity.  Thus, although statistical tests using click probabilities cannot practically distinguish a Fock state from a coherent state, homodyne detection does achieve this. 

We have also found that there is unsuppressed noise of this kind in a wide class of states that combine thermal excitation, displacement, and squeezing operations.  The predicted magnitude of the noise depends non-trivially upon the parameters describing those operation, which is to say that it allows us to accessing new information about the quantum state of the radiation field, that differentiates among different possibilities, even for large occupation numbers.  

Whereas homodyne detection involves a local oscillator operating at the same frequency as the detector, in heterodyne detection the local oscillator operates at a different frequency.  Heterodyne detection potentially accesses still more information, and can be analyzed along similar lines, but we won't do that here.  


\bigskip

{\it Sources of acoherence}

\bigskip

As was previously mentioned, and as is well known, linear coupling to a classical source produces radiation fields that satisfy the coherent state hypothesis.  This plausibly applies to most weak-field sources of astrophysical gravitational radiation, for which a linear expansion of the gravitation field around a stationary (non-radiating) solution is accurate.  

Also well known, although less straightforward, is that there are circumstances in which gravitational radiation is predicted to be approximately thermal, and thus definitely does {\it not\/} satisfy the coherent state hypothesis.  Notable examples are Hawking radiation from black holes and gravitation radiation associated with inflation.  These are typically calculated using methods that bring in quantum theory from an early stage.  (Though they can occur at weak coupling, and in fixed external fields, they generally bring in tunneling or effective non-locality.) 

Here we will mention three plausible sources of deviations from the coherent state hypothesis -- i.e., {\it acoherence\/} -- that arise in the context of astrophysical sources that are now under active observation in gravitational wave astronomy.  

{\it Overtones}: In the aftermath of a black-hole merger, just before the merged object settles into a stationary black hole, there is a ``ring-down'' phase characterized by small -- but not minuscule! -- oscillations around the stationary solution.  

Here Einstein's equations expanded to first order in the metric give very precise predictions for the spectrum of quasinormal modes that are well understood~\cite{ReggeWheeler}.  
Interestingly, the perturbative approach can be continued.  Taken to the next order, it gives predictions for the production of quasi-normal mode overtones~\cite{QNM1,QNM2,QQNM}. With the $l=2,m=2$ even parity mode as the fundamental mode, second harmonic conversion occurs with amplitude $\frac{|A_{4,4}^{(2)}(2\omega)|}{|A_{2,2}^{(1)}(\omega)|^2}\sim 0.15$ at the source.  Non-perturbative calculations in numerical general relativity~\cite{NumericalGR1,NumericalGR2} bear out this picture, which suggests that the overtones are observable (though not an easy target!).  

In quantum optics, it is well known that second harmonic generation results in squeezing, especially in the fundamental mode, of a sub-Poissonian nature~\cite{Mandel}. It is a process where annihilation of two quanta in the fundamental mode produce one frequency up-converted quantum in the second harmonic mode. The significant conversion efficiencies relevant to the gravitational context described above suggest that in an effective quantum description, the observed gravitational quasi-normal modes are necessarily squeezed. Since the real part of their frequencies can be in the kilo-Hz range, and amplitudes that LIGO can access in parallel, it is not inconceivable that bar detectors equipped with phase sensitive measurement strategies will be able to observe the excess phase noise from the squeezing, though click detectors will not observe excess noise.  

From a theoretical perspective, the overtones generated during ring-down provide a beautiful laboratory for controlled exploration of non-linear sources and the deviations from the coherent state hypothesis that can result.  The corresponding experiments are far from easy, but perhaps not terminally unrealistic.

{\it Subharmonics}: The phenomena known as pair creation in particle physics, parametric down-conversion in quantum optics, and Bogoliubov transformations in other contexts, all stem from the same mathematical source, namely an effective Hamiltonian of the schematic form 
\begin{equation}
H_{\rm eff.} ~ \sim J \, (a^\dagger_+ a^\dagger_- \, + a_{+} a_{-} )
\end{equation}
where $J$ is the source and in the simplest cases the creation operators correspond to creation of particles moving in opposite directions.  In different applications there will be polarization indices, a non-trivial superposition over angles, and other complications.  Since gravity couples non-linearly, it supports pair creation of this kind, wherein $J$ can be a time-dependent, effectively classical macroscopic gravitational field.  Hamiltonians of this kind create squeezed states directly.  

Related to this, in the context of quasi-normal modes, frequency down-conversion is possible, even.  Here, when the merger dynamics excites the second harmonic, parametric down conversion into the fundamental mode squeezes the fundamental mode. 

{\it Exploratory - transients}: Prior to the ringdown, at the (ill-defined) onset of the merger process, there is a period of intensely non-linear gravitational interaction.  While it is difficult to compute analytically, the gravitational radiation emitted during this period is far removed from from the ``linear coupling to a classical source'' paradigm, and seems not unlikely to involve significant deviation from the coherent state hypothesis.  Here is where tests of the kind proposed above can offer new windows into strong-field gravitational dynamics, and pose new challenges for both theorists and observers.

\end{document}